\documentclass[iop]{emulateapj}

\usepackage{amsmath}

\newcommand{\thetabold}{\mbox{\boldmath$\theta$}}

\begin{document}

\title{On the inversion of Stokes profiles with local stray-light contamination}
\author{A. Asensio Ramos \& R. Manso Sainz}
\affil{Instituto de Astrof\'{\i}sica de Canarias, 38205, La Laguna, Tenerife, Spain \\
Departamento de Astrof\'{\i}sica, Universidad de La Laguna, E-38205 La Laguna, Tenerife, Spain \email{aasensio@iac.es}}

\begin{abstract}
Obtaining the magnetic properties of non-resolved structures in the solar photosphere is 
always challenging and problems arise because the inversion is carried out through the numerical
minimization of a merit function that depends on the proposed model. We investigate the reliability of inversions in which
the stray-light contamination is obtained from the same observations as a local average. In this case,
we show that it is fundamental to include the covariance between the observed Stokes 
profiles and the stray-light contamination. The ensuing modified merit function of the
inversion process penalizes large
stray-light contaminations simply because of the presence of positive correlations between the observables
and the stray-light, fundamentally produced by spatially variable systematics. We caution that using 
the wrong merit function, artificially large stray-light 
contaminations might be inferred. Since this effect disappears if the stray-light contamination
is obtained as an average over the full field-of-view, we recommend to take into
account stray-light contamination using a global approach.
\end{abstract}

\keywords{methods: data analysis, statistical --- techniques: polarimetric --- Sun: photosphere}

\maketitle

\section{Introduction}
The quantitative investigation of the magnetism of structures in the solar atmosphere
is done through the analysis of the observed Stokes profiles.
Particularly difficult is to infer the properties of weakly magnetized regions of the
solar surface such as the quiet internetwork.
The reason is that the observed polarization signals often stay at the detection limit
of modern spectro-polarimeters because the spatial resolution is still not
high enough to resolve the smallest magnetic structures. 
Therefore, the magnetic properties are inferred assuming unresolved structures,
which is model dependent.

Due to lack of information, it is customary to apply relatively simple
models to explain the observations. The model parameters are estimated
using a least-squares (or, equivalently, maximum-likelihood) approach
\citep[e.g.,][and many others]{skumanich_lites87,sir92,socas_trujillo_ruiz00,frutiger00,lagg04,asensio_trujillo_hazel08} or 
using a fully Bayesian approach \citep{asensio_martinez_rubino07,asensio_hinode09}.
For the inversion of unresolved structures, the idea of using several components
that contribute to the observed Stokes profiles has been around since its introduction by
\cite{lites_skumanich90}. In its simplest form,
the method consists in using a filling factor $\alpha$ of the pixel that accounts for 
a non-magnetized component (either stray-light or a pure non-magnetic plasma) 
and the remaining $1-\alpha$ fraction of the pixel is filled 
by a magnetic component.

This paper points out that, in case the stray-light contamination is obtained
directly from the observations as an average over a given field-of-view, 
some modifications are necessary in the inversion procedure to take into
account the eventual presence of correlation between the observed Stokes profiles and the
stray-light. This modification has the remarkable property of penalizing
large stray-light contaminations.

\section{Discussion}
In order to obtain information on the magnetic and thermodynamical
properties of the solar atmosphere, one proposes a model atmosphere
that depends on a set of parameters \thetabold\
(e.g., the temperature $T$ at one or several heights, hydrogen density, 
magnetic field strength and inclination, etc). This model is
used to synthesize the Stokes vector $\mathbf{O}^\mathrm{mod}(x, y, \lambda; \thetabold)$ for an arbitrary
number of spectral lines at given spatial position $(x,y)$ and wavelength $\lambda$. Expressing the observed
Stokes profiles as $\mathbf{O}^\mathrm{obs}(x,y,\lambda)$, 
it is customary to obtain the ``best'' parameters $\widehat{\thetabold}$ as
those minimizing the 
following merit function \cite[e.g.,][]{numerical_recipes86}:
\begin{equation}
\chi^2(x,y) = \frac{1}{4N}\sum_{i=1}^4 \sum_{j=1}^{N} \frac{\left[O_i^\mathrm{mod}(x,y,\lambda_j,\thetabold) - O_i^\mathrm{obs}(x,y,\lambda_j)\right]^2}{\sigma^2_i(x,y,\lambda_j)},
\label{eq:chi2}
\end{equation}
where the sum over $j$ is extended to all $N$ observed wavelength points.
In this equation, $\sigma_i^2(x, y, \lambda_j)$ represents the variance of the numerator,
for each wavelength $\lambda_j$, Stokes parameter
$i=I, Q, U$, and $V$, and at each location $(x, y)$, due to possible uncertainties in the observations (measurement errors and noise). 
This is different from ``real'' variations due to space-time fluctuations (e.g., intensity contrast due to granule-intergranule fluctuations).
Usually, the model is known with certainty 
---as when we fit a Gaussian profile to an observed spectral line---,
and $\sigma_i^2(x, y, \lambda_j)$ is just the noise variance $\sigma_n^2$ of the data.
This noise variance can be estimated from the observations, ideally, by taking
several observations of the same object under identical observational conditions.
Unfortunately, this is often just not possible. It is then customary
in spectroscopic observations to select a continuum window and, assuming that it
should be spectrally flat, all fluctuations are due to noise. The flatness assumption
is usually not fulfilled due to systematic effects and the estimated variance
might be larger than the one associated with random noise effects.\footnote{We 
shall not consider here the expected variation of $\sigma_n^2$ with wavelength 
across strong spectral lines. This is a very interesting problem 
that will be considered elsewhere.}

\begin{figure}[!t]
\resizebox{\hsize}{!}{\includegraphics{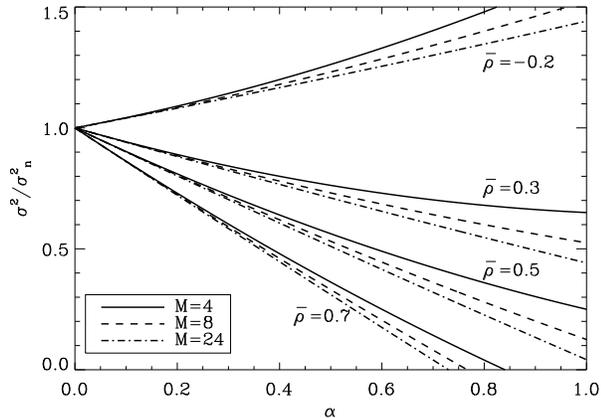}}
\caption{Value of the variance in units of $\sigma^2_n$ as defined
in Eq. (\ref{eq:sigma2_final}) for 
different values of $M$ and for four different values of $\bar{\rho}$,
the covariance between the observed profile and the stray-light. Note that
$\bar{\rho}$ depends on $M$ but, for simplicity, we have assumed that 
$\mathrm{Cov} \left[O_i^\mathrm{obs}(x,y),O_i^\mathrm{obs}(x',y') \right]$
is constant, so that $\bar{\rho}$ is independent of $M$.
Although not shown to avoid crowding, when the covariance is maximum 
($\bar{\rho}=1$), values of the stray-light contamination above 
$\sim$1/2 are strictly forbidden. The exact value of this limit
can be obtained from Eq. (\ref{eq:alpha_max}).\label{fig:variances}}
\end{figure}


However, it is also possible that the model contribute to the variance uncertainty
by incorporating explicitly some observable ---for example, if we fit a Gaussian profile
to an observed spectral line, but keeping the equivalent width of the observed
line profile. We study here an example of this type that arises naturally when
considering stray light in spectropolarimetric observations.

The proper interpretation and inversion of spectropolarimetric observations with
spatial resolution requires some treatment of stray light.
``Stray light'' refers to the unavoidable spread of light from different regions
on the source. It is an effect produced by the
extended tails of the telescope's point spread function, which causes that
a significant fraction of the photons detected in one pixel results from regions
outside the Airy disk.
As a consequence, there is always some amount of contamination from regions other
than the region that was supposed to be imaged by the instrument.
This problem has become especially pressing recently, with the advent of
high-resolution space-borne spectropolarimeters like the SP
\citep[][]{lites_hinode01} aboard \emph{Hinode} \citep{kosugi_hinode07}.
To deal with this problem when carrying out inversions of Stokes profiles from
quiet regions of the solar photosphere, the following model has been proposed
\citep[e.g.,][]{lites_skumanich90,orozco_mhd07,orozco_hinode07}:
\begin{equation}
\mathbf{O}^\mathrm{mod}(x,y;\thetabold) = \alpha \mathbf{D}(x,y) + (1-\alpha) \mathbf{S}(x,y;\thetabold),
\label{eq:model_stray}
\end{equation}
where $\mathbf{S}(x,y,\thetabold)$ are the Stokes profiles emerging
from a model atmosphere representative for a magnetized region
occupying a fraction $1-\alpha$ of the pixel (from now on, we drop the dependence
of the variables on $\lambda$). The term 
$\mathbf{D}(x,y)$ is a stray-light contamination profile that is obtained from a
local average of the $M$ pixels around the pixel of interest (of the order of 1 arcsec$^2$
around the pixel of interest for Hinode observations):
\begin{equation}
\mathbf{D}(x,y) = M^{-1} \sum_{(x',y') \in \Omega} \mathbf{O}^\mathrm{obs}(x',y'),
\end{equation}
where $\Omega$ is the set of $M$ pixels that is considered to be affecting the pixel
of interest. The parameter $\alpha$ quantifies the amount of stray-light contamination.
In this case, the denominator of the $\chi^2$ function in Eq. (\ref{eq:chi2}) is:
\begin{equation}
\sigma^2_i(x,y) = \mathrm{Var} \Big[ \alpha D_i(x,y)+(1-\alpha) S_i(x,y) - O_i^\mathrm{obs}(x,y) \Big].
\end{equation}

The presence of $D_i(x,y)$, which depends on the observations $O_i^\mathrm{obs}(x',y')$, produces that
the variance is not just the noise variance because a certain degree of correlation
might exist between $O_i^\mathrm{obs}(x,y)$ and $D_i(x,y)$. The reason is that both quantities are
affected by noise and systematic effects and they are the result of a complex reduction
process that might introduce some correlation. A straightforward calculation allows us to simplify the
variance to:
\begin{equation}
\sigma^2_i(x,y) = \left( 1+ \frac{\alpha^2}{M} \right) \sigma_n^2 - 2\alpha \mathrm{Cov} \left[O_i^\mathrm{obs}(x,y),D_i(x,y) \right].
\label{eq:variance_final}
\end{equation}

It is of interest to define ${\bar{\rho}}$ as the covariance measured in units of the noise variance:
%
\begin{eqnarray}
\bar{\rho} \sigma^2_n &=& \mathrm{Cov} \left[O_i^\mathrm{obs}(x,y),D_i(x,y) \right] \nonumber \\
&=& \frac{1}{M} \sum_{(x',y') \in \Omega} \mathrm{Cov} \left[O_i^\mathrm{obs}(x,y),O_i^\mathrm{obs}(x',y') \right],
\label{eq:f_definition}
\end{eqnarray}
%
which explicitly shows the dependence of $D_i(x,y)$ on the \emph{observed} Stokes
profiles and also indicates that $\bar{\rho}$ behaves like an average
correlation coefficient. Plugging this expression into Eq. (\ref{eq:variance_final}) results in:
\begin{equation}
\sigma^2_i(x,y) = \left( 1+ \frac{\alpha^2}{M} -2\alpha \bar{\rho} \right) \sigma_n^2.
\label{eq:sigma2_final}
\end{equation}


\begin{figure*}
\centering
\includegraphics[width=0.5\textwidth]{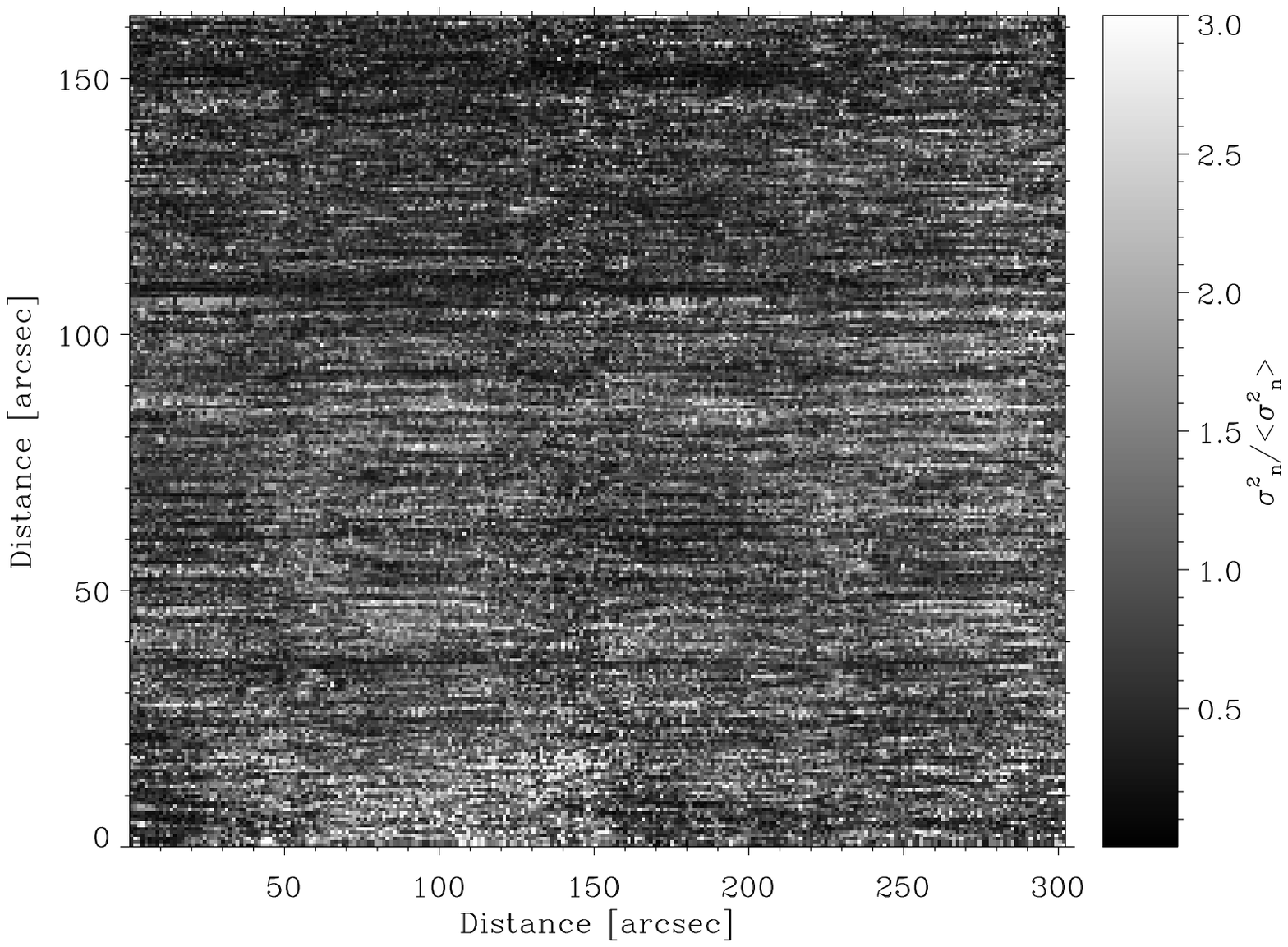}%
\includegraphics[width=0.5\textwidth]{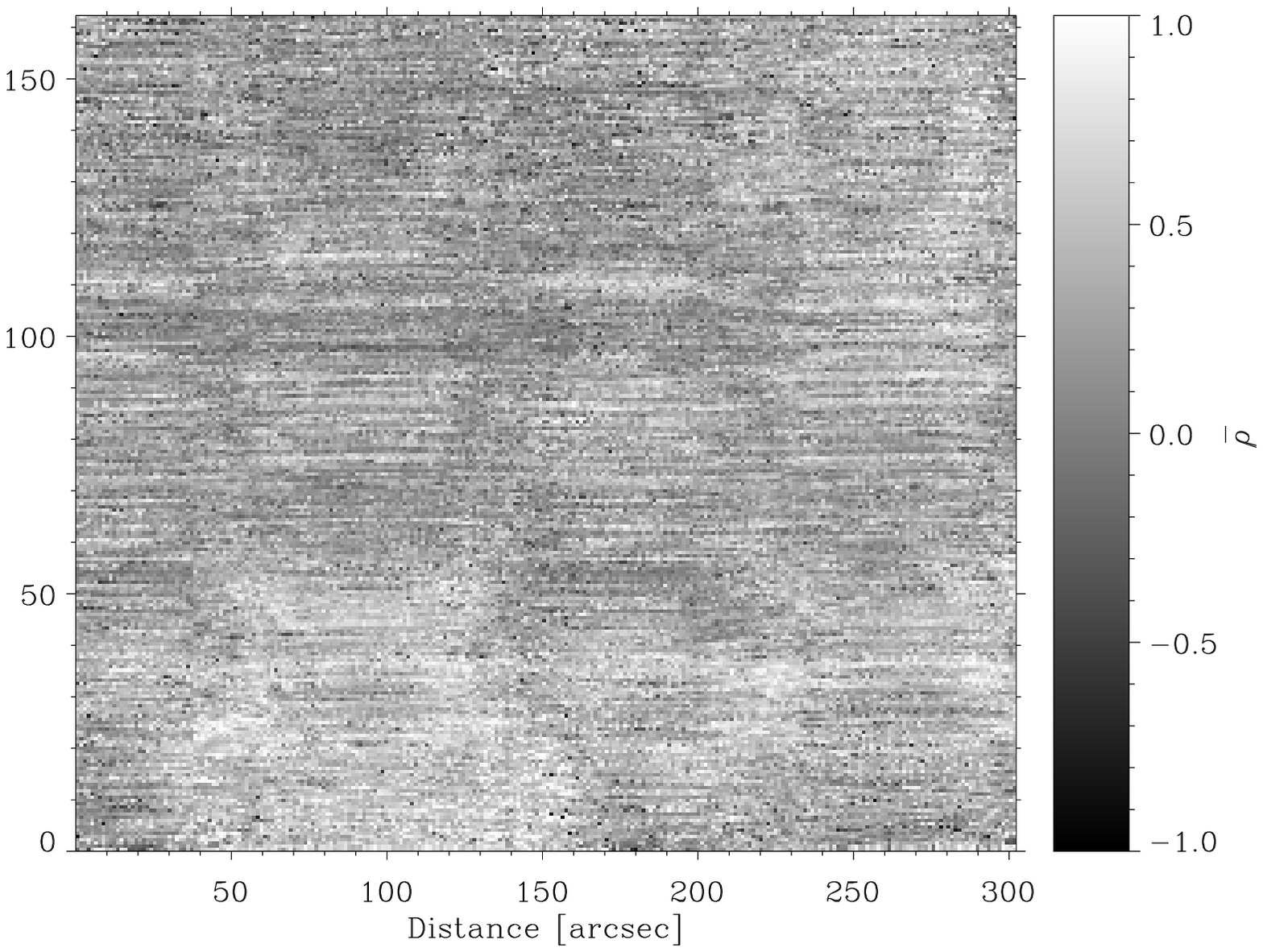}
\includegraphics[width=0.5\textwidth]{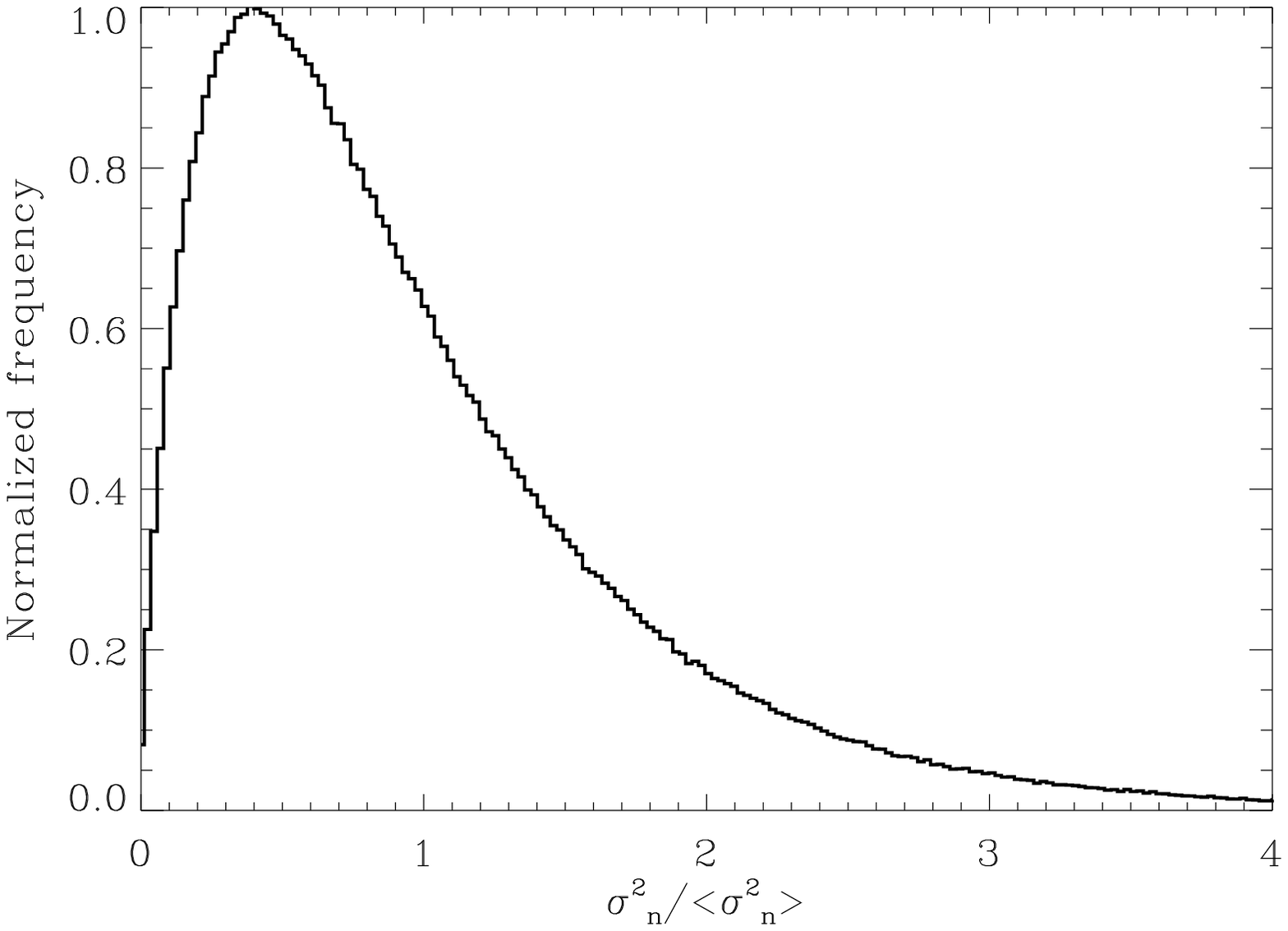}%
\includegraphics[width=0.5\textwidth]{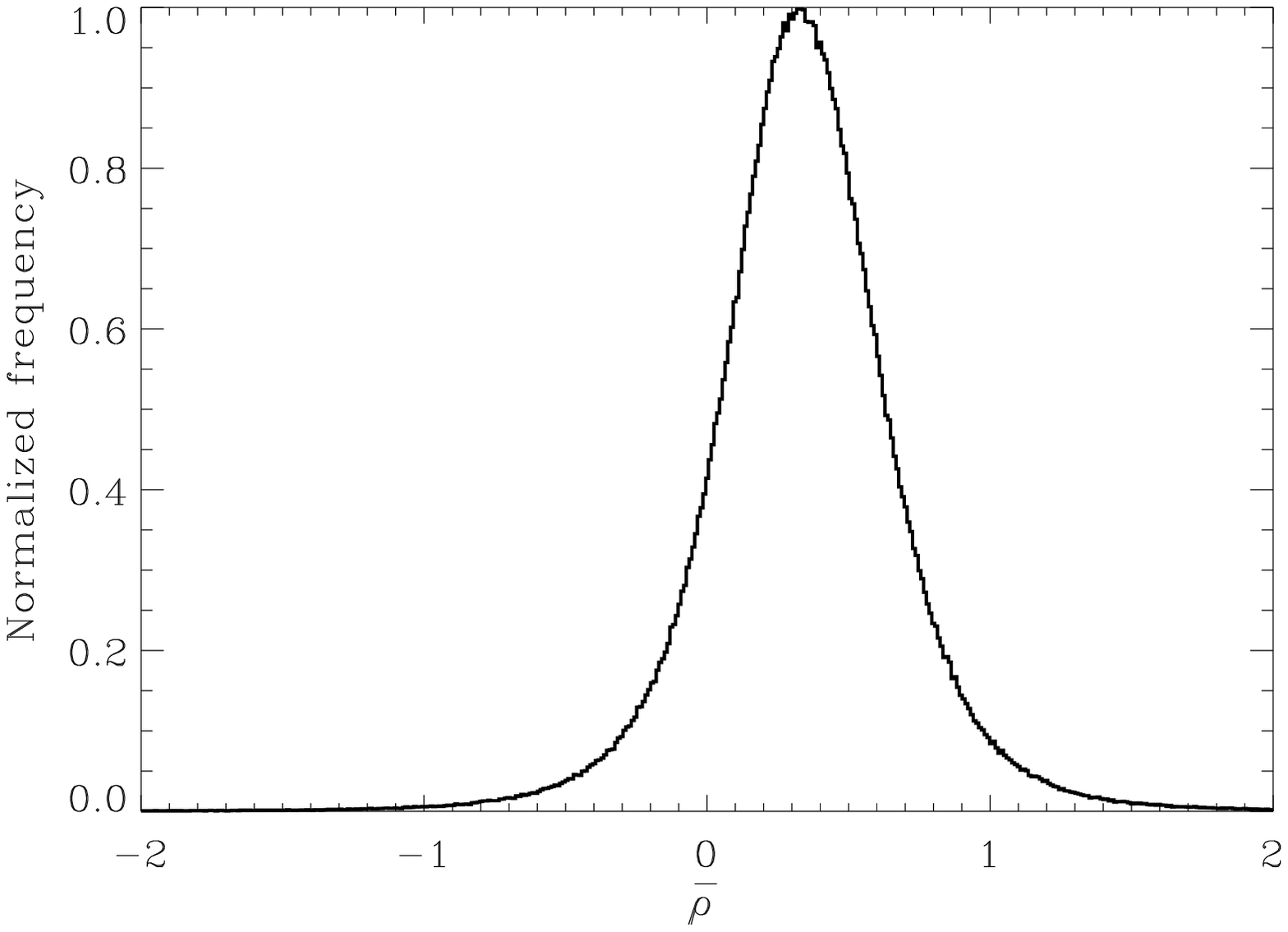}%
\caption{Spatial variation of the variance (upper left panel) and covariance between the local profile and the
stray-light contamination using a local environment of 1.5$"$ around each pixel (upper right panel). The
Hinode observations are those used by \cite{lites08}. This figure is representative of
a window in the continuum and it is expected to change if other points in the spectral
direction are used because the systematic effects might change. The lower
panels show the histogram of each image. Note that $|\rho|>1$ are found
because of the small sample with which it is calculated, thus leading
to statistical fluctuations in which the covariance and the variance
are not strictly compatible.\label{fig:data_covariance}}
\end{figure*}

This expression for the variance has several interesting properties. It depends
on the stray-light contamination coefficient $\alpha$. The term in $\alpha^2$
accounts for the inclusion of additional noise variance coming from the stray-light
profile. Its influence is heavily reduced when averaging over many pixels, decreasing 
in proportion to the number of pixels added. The linear term
depends on the covariance between the observed Stokes profile at a given
wavelength and the average profile that is considered as stray-light contamination. 
This term also goes to zero when many pixels are considered for the averaging, so
we recommend to obtain the stray-light contamination profile as an average over the
full field-of-view.
A non-zero contribution can be produced by any spatially variable systematic effect present
in the final focal plane produced by the camera (flatfield systematics or fringes) 
or by any optics before (fringes). Additionally, corrections carried out 
during data reduction can introduce correlations between surrounding
pixels. Finally, data compression like the JPEG compression used 
by Hinode \citep{lites_jpeg02} to optimize telemetry can also
introduce correlation artifacts. As a consequence of this dependence on $\alpha$, the $\chi^2$ merit 
function to be optimized to get the maximum-likelihood parameters $\widehat{\thetabold}$
is different from the one commonly used in the literature\footnote{Note that this modifies the 
likelihood function used by \cite{asensio_hinode09} in the Bayesian
framework too.}. We shall see shortly that this can affect the results significantly.

Although $-1\le \bar{\rho}\le 1$, strong anticorrelations ($\bar{\rho}<0$) 
can be effectively discarded in our case.
Figure~\ref{fig:variances} shows how $\sigma_i^2/\sigma_n^2$ [see Eq. (\ref{eq:sigma2_final})]
varies with $M$ and $\bar{\rho}$, assuming that $\mathrm{Cov} \left[O_i^\mathrm{obs}(x,y),O_i^\mathrm{obs}(x',y') \right]$
is constant on $\Omega$. 
Taking $M=24$ is roughly equivalent to consider a neighborhood $\sim$1.5$"$ around the
pixel of interest at Hinode's spatial sampling. 
As seen in Fig. \ref{fig:variances}, the dependence of the variance on $M$ is 
relatively weak and the parabolic 
shape of the curves tends to a straight line as $M$ increases.
If $\bar{\rho}>0$, stray-light contaminations above 
the following limit are forbidden:
\begin{equation}
\alpha_\mathrm{max}=\bar{\rho} M-\sqrt{(\bar{\rho}M)^2-M},
\label{eq:alpha_max}
\end{equation}
which has been obtained as the solution of $\sigma^2_i(x,y)=0$.
This quantity rapidly tends to $(2\bar{\rho})^{-1}$ for increasing values of $M$ and
is smaller than 1, whenever $\bar{\rho}>1/2$ and $M \geq (2\bar{\rho}-1)^{-1}$. If $\bar{\rho}<0$, the
behavior is the opposite and large values of the stray-light contamination
are favored.

It is important to estimate the covariance between stray-light
contamination and observed profiles. To this end, we use the
quiet Sun map observed by \cite{lites08} to compute the
stray-light contamination in a typical Hinode observation 
as an average over a window of $\sim$1.5$"$
around every pixel of interest. 
The upper left panel of Fig. \ref{fig:data_covariance} shows
the variance of the observed Stokes $I$ continuum 
normalized to the average variance of the whole map. This is equivalent to 
the noise variance in case the continuum is assumed to be spectrally flat. Our
experience is that the average value of the standard deviation for Stokes $I$ is a factor 3-4 larger than the
noise estimation in Stokes $Q$, $U$ and $V$ made by \cite{lites08}. We
assign this difference to the unavoidable presence of systematic effects 
in Stokes $I$ due to flatfielding procedures. Since
they change from pixel to pixel, they can be absorbed as part
of the noise, though probably not normally distributed. 
Note the presence of conspicuous horizontal
stripes that show the presence of pixels along the slit and at the
points of the camera associated with continuum wavelengths which present a somewhat 
higher variance (higher systematic effects).
The upper right panel of Fig. \ref{fig:data_covariance} shows the value of
$\bar{\rho}$ for the whole map, which has been calculated using a
small window in the continuum from 6302.915 \AA\ to 6303.002 \AA. Strictly speaking, the covariance should
have been calculated using many realizations of the measurement process with the underlying
solar profile fixed. Assuming some kind of {\em ergodicity}, we estimated the
covariance using a wavelength window in the continuum. For consistency, we
also estimated the covariance along the time using the time 
series of \citep{lites08}, with very similar results. 
Again, the presence of horizontal stripes
is clear, pointing to the systematic character of such defects. 
On average, $\bar{\rho} \sim 0.45$, although the distribution of covariances
is clearly heavy-tailed, with the presence of large values of
$\bar{\rho}$ much more frequently than in the case of a Gaussian distribution.
The lower panels of Fig. \ref{fig:data_covariance} present histograms of the
maps. A characteristic is that it is possible to find unphysical $|\rho|>1$, a consequence
of the small sample with which this quantity is computed.

\begin{figure}
\centering
\includegraphics[width=\columnwidth]{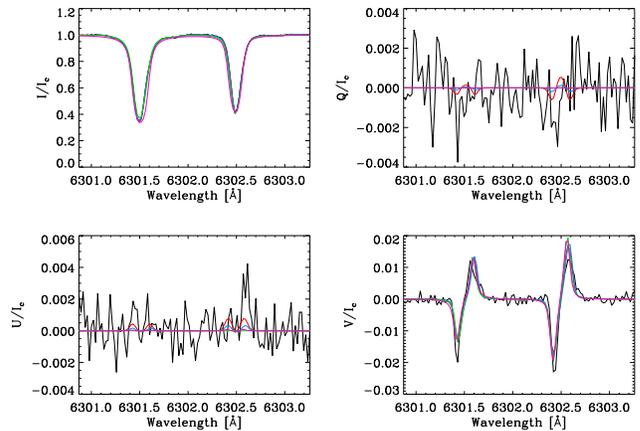}
\caption{Observed Stokes profiles (black solid line) of a pixel in the field-of-view
observed by \cite{lites08}. Inversions with $\bar{\rho}=0$ are shown in red, $\bar{\rho}=0.5$ in blue, $\bar{\rho}=0.7$ in
green and $\bar{\rho}=1$ in magenta. All of them give reasonably good results but the
model parameters change dramatically as shown in Tab. \ref{tab:results_MAP}.
Note that the cases of large $\bar{\rho}$ are not realistic in this situation since we adopt
a covariance that is too large with respect to the correct one.\label{fig:stokes}}
\end{figure}

It is customary in standard inversion codes to introduce different weights for
each Stokes parameter so that Stokes $I$ does not dominate the merit function. 
One may think that this alleviates the effect of covariance
on the inferred parameters. However, the inclusion of
weights does not mimic properly the form of $\chi^2$
because the variance term in Eq. (\ref{eq:sigma2_final}) depends explicitly on one of the model parameters,
although the intuition is partially right.
Therefore, one would expect differences 
with respect to the parameters obtained 
with a standard $\chi^2$. This is
indeed the case, as shown in Tab. \ref{tab:results_MAP}. This table shows
the results of a least-squares fit to the Stokes profiles shown in
Fig. \ref{fig:stokes} for different values of $\bar{\rho}$. The model proposed is
that of Eq. (\ref{eq:model_stray}) where the magnetic component is a 
Milne-Eddington atmosphere and the stray-light contamination is obtained from
a local average of $M=24$ pixels around the one of interest. The table presents
the magnetic field strength $B$ and inclination $\theta_B$, together with the stray-light
contamination $\alpha$ that minimizes the $\chi^2$ for different
values of $\bar{\rho}$ and the value of the reduced $\chi^2$ at the
minimum. A striking effect of the inclusion of stray-light
is that the value of $\alpha$ decreases as soon as $\bar{\rho}$ increases,
a direct consequence of the presence of positive correlations between the observed
profile and the stray-light. Therefore, the magnetic field strength has
to be adjusted accordingly in order to maintain the magnetic flux density, much in the direction of what has
been explored by \cite{martinez_gonzalez06}. Note that the cases with $\bar{\rho}>0.5$
are not so common in the observations. As a consequence, since such
large values are not representative of the sample profile, the fits are not good and they
should be taken as indicative of what would happen in such a
limiting case. Finally, it is important to point out that this behavior
would be reduced if the field is strong enough to produce a substantial Zeeman
splitting because the magnetic field strength can be inferred directly
from the splitting.

\section{Conclusions}
We have demonstrated that the inversion of Stokes profiles with models that 
include stray-light contaminations obtained from the same observations has to 
be carried out with care. In general, the $\chi^2$ merit function is modified
and has to include the effect of the variance of the stray-light profile and 
the covariance between such profile and the one being inverted. The most important 
consequence is that the noise variance now depends on the stray-light contamination 
parameter, $\alpha$.

A first contribution adds quadratically with the noise variance and is inversely
proportional to the number of pixels that contribute to the stray-light profile.
When the number of pixels is large enough, this term turns out to be almost
negligible. The fundamental reason is that the signal-to-noise ratio of the
stray-light profile increases and becomes much larger than that of the local
profile, which dominates then the variance. If the stray-light
contamination profile is obtained as an average over the full field-of-view,
the effect of correlation diminishes considerably and the modified $\chi^2$
converges to the standard $\chi^2$ used in the past.

\begin{table}
\caption{Maximum-likelihood parameters}              
\label{tab:results_MAP}      
\centering                                      
\begin{tabular}{c c c c c}          
\hline\hline
$\bar{\rho}$ & $\widehat{B}$ & $\widehat{\theta}_B$ & $\widehat{\alpha}$ & $\chi^2_\mathrm{min}$\\    
\hline                                   
    0.0 & 844 & 155 & 0.91 & 2.0\\      
    0.5 & 558 & 162 & 0.88 & 3.8\\
    0.7 & 114 & 176 & 0.46 & 11.3\\
    1.0 & 62 & 169  & 0.02 & 13.8\\
\hline                                             
\end{tabular}
\end{table}

A second contribution adds linearly and is much more delicate. It accounts for
all the systematic effects (fringes, flatfield systematics, data reduction effects, etc.) that plague the
observations. The presence of systematics that are well above
the noise level make both profiles vary similarly. The covariance term takes this
into account by disfavoring models with large stray-light contaminations. In the
absence of this term, any inversion method prefers to use a large stray-light
contamination because the profiles become much more similar using the standard and
incorrect $\chi^2$. A simulated exercise have shown that neglecting the presence
of this covariance might lead to artificially strong magnetic field strengths and 
large stray-light contaminations.

\begin{acknowledgements}
We thank C. Beck, M. J. Mart\'{\i}nez Gonz\'alez and J. A. Rubi\~no Mart\'{\i}n for useful suggestions. Financial support by the 
Spanish Ministry of Science and Innovation through projects AYA2010-18029 (Solar Magnetism and Astrophysical 
Spectropolarimetry) and Consolider-Ingenio 2010 CSD2009-00038 is gratefully acknowledged.
\end{acknowledgements}



\end{document}